\documentclass[preprint,superscriptaddress,nofootinbib,12pt]{revtex4-1}
\usepackage{amssymb, amsmath, graphicx}
\usepackage{color}

\pdfoutput=1

\def\lag{\mathcal{L}}
\def\lagint{\lag_\text{int}}
\def\op{\mathcal{O}}

\def\chibar{\bar{\chi}}

\def\beq{\begin{equation}}
\def\eeq{\end{equation}}

\def\bit{\begin{itemize}}
\def\eit{\end{itemize}}
\def\ben{\begin{enumerate}}
\def\een{\end{enumerate}}

\begin{document}

\title{Light Dark Matter Anomalies After LUX}

\author{Moira I. Gresham}
\affiliation{Whitman College, Walla Walla, WA 99362}
\author{Kathryn M. Zurek}
\affiliation{Michigan Center for Theoretical Physics, University of Michigan, Ann Arbor, MI 48109, USA}

\begin{abstract}

We examine the consistency of light dark matter (DM) elastic scattering in CoGeNT, DAMA, and CDMS-Silicon in light of constraints from XENON, CDMS, LUX, PICASSO and COUPP.  We consider a variety of operators that have been employed to reconcile anomalies with constraints, including anapole, magnetic dipole, momentum-dependent, and isospin-violating DM.  We find that elastic scattering through these alternative operators does not substantially reduce the tension between the signals and the null constraints for operators where at least two of the three purported signals map onto a common space in the DM mass--scattering cross-section plane.  Taking a choice of the scintillation efficiency that lies at the $-1 \sigma$ region of the Manzur et al measurement relieves tension between signals and the LUX constraint---in particular for a magnetic dipole interaction and a xenophobic interaction (though for the latter the signal regions do not substantially overlap).  We also find that modest changes in the halo model does not alter this result.  We conclude that, even relaxing the assumption about the type of elastic scattering interaction and taking a conservative choice for the scintillation efficiency, LUX and the results from other null experiments remain in tension with a light DM elastic scattering explanation of direct detection anomalies. 
\end{abstract}
\preprint{MCTP-13-37}

\maketitle
\tableofcontents

\section{Introduction}

Low mass dark matter (DM) anomalies have, to this point, shown remarkable resilience to experimental constraints.  Old anomalies have remained and new ones appeared, all while new constraints have continued to close the allowed parameter space for an elastically scattering light DM particle in the 7-12 GeV mass window that can explain the signals.  The first anomaly appeared from DAMA, which reported a high significance modulation consistent with light DM recoiling off Sodium Iodide crystals \cite{Bernabei:2008yi,Bernabei:2010mq}.  The CoGeNT experiment subsequently reported an excess of events at low energy consistent with light DM scattering off Germanium \cite{Aalseth:2010vx}; later it was found that approximately half of these events were from surface contamination \cite{Aalseth:2012if}.   

These anomalies became the target for searches of light DM, and the null results from XENON10, XENON100, PICASSO, COUPP, CDMS germanium low energy and CDMSLite constrained the region \cite{Angle:2011th,Aprile:2012nq,Archambault:2012pm,Behnke:2012ys,Ahmed:2010wy,Agnese:2013lua}.  The strongest constraints are derived from XENON in the spin-independent case, though the robustness of these limits is subject to nuclear recoil energy calibration uncertainties near threshold, encapsulated in the ${\cal L}_\text{eff}$ and $Q_y$ parameters (see, for example, \cite{Sorensen:2010hq,Sorensen:2012ts} for a discussion) which made their constraints controversial.  In the end, however, these constraints appeared to be so strong that even non-standard models of WIMP DM scattering ({\em e.g.} \cite{Chang:2009yt,Fitzpatrick:2010br,Kopp:2011yr,DelNobile:2012tx,DelNobile:2013cta,Boehm:2013qva,Buckley:2013gjo,Frandsen:2013cna}) did not evade the constraints.  Implementing different velocity distributions also did not relieve the tension \cite{Fitzpatrick:2010br,Fox:2010bz,Gondolo:2012rs,DelNobile:2013cta}.

Interest revived, however, when CDMS reported an excess of three events in Silicon data at threshold consistent with a light DM candidate \cite{Agnese:2013rvf}.  The preferred region is also naively consistent with the CoGeNT excess, though again marginally in conflict with the XENON constraint.  Since the targets in DAMA, CoGeNT, and CDMS are different than in XENON, the constraints may not be compared in a model-independent fashion.  For example, an effort to tune away the XENON constraint via isospin violation, which reduces the DM scattering cross-section off of Xenon, can successfully reduce the tension (though the tension with the CDMS germanium and CDMSLite results remains) \cite{Feng:2011vu}.

Most recently, LUX has weighed in on the light DM fray with a low nuclear recoil energy constraint, their result \cite{Akerib:2013tjd} reaching to a nuclear recoil threshold of 3 keV.  For an interpretation of the CDMS three events with spin-independent scattering, with equal DM coupling to the proton and neutron, at a cross-section $2 \times 10^{-41} \mbox{ cm}^2$, LUX would see approximately 1500 events.  Given the presence of few electron recoil events leaking into the nuclear recoil band, LUX is able to put a strong constraint on the entire preferred region of the CDMS-Silicon three events.

The purpose of the present paper is to project the LUX, as well as XENON10, XENON100, CDMS germanium low energy, CDMSLite, COUPP and PICASSO constraints onto the space for scattering through standard and non-standard types of interactions, looking beyond the usual spin-independent and -dependent scattering operators.  In many models of DM, the leading interactions may be momentum (or velocity) dependent \cite{Bagnasco:1993st,Sigurdson:2004zp,Chang:2009yt,Feldstein:2009np,Feldstein:2009tr,Banks:2010eh,An:2010kc,Barger:2010gv,Belanger:2013tla}.  The simplest cases to consider are interactions through the DM anapole and dipole operators \cite{Fitzpatrick:2010br}, or through pseudoscalars.   
In particular, the operators we consider are
\begin{eqnarray}
{\cal O}_a & = & \bar{\chi} \gamma^\mu \gamma_5 \chi {A'}_\mu \label{eq: anapole op}\\
{\cal O}_d & = & \bar{\chi} \sigma^{\mu \nu} \chi {F^{(')}}_{\mu \nu} \label{eq; mag dipole op}\\
{\cal O}_\phi & = & \bar{\chi}(a + b \gamma_5) \chi \phi. \label{eq: scalar mediator op}
\end{eqnarray}
The first two operators are the anapole and dipole, respectively.  The gauge field may or may not be the Standard Model (SM) $U(1)$ in the dipole case. One attractive scenario arises when a dark gauge field ($A'$) mixes with hypercharge.  The anapole is attractive because it is the leading operator through which Majorana DM can couple to the nucleus through a vector interaction.  The dipole couples the DM spin to the field strength, and naturally arises in some models of composite DM \cite{Banks:2010eh,Bagnasco:1993st}. 
Given an effective nucleon interaction of the form $\phi \bar{N}(c + d \gamma_5)N$, the following effective operators are generated when $\phi$ is integrated out:
\begin{align}
\op_1&=\bar{\chi} \gamma^5 \chi \bar{N} N \label{eq: q2SI op}\\
\op_2&=\bar{\chi}  \chi \bar{N}\gamma^5 N \label{eq: q2SD op} \\
\op_3&=\bar{\chi}\gamma^5  \chi \bar{N}\gamma^5 N \label{eq: q4SD op}
\end{align} in addition to the standard spin-independent operator $\bar{\chi} \chi \bar{N} N$. If $a=0$ and/or $c=0$, say, for symmetry reasons, then the standard spin-independent operator is absent. The operators of Eqs.~\eqref{eq: q2SI op}-\eqref{eq: q4SD op} were highlighted in \cite{Chang:2009yt} as leading to WIMP-nucleus interactions with leading $q^2$ dependence. $\op_1$ leads to a $q^2$-suppressed, spin-independent interaction; $\op_2$ to a $q^2$-suppressed, spin-dependent interaction; and $\op_3$ to a $q^4$-suppressed, spin-dependent interaction. 

On the other hand, the momentum and velocity dependence for the anapole and magnetic dipole operators are more subtle, and depend on the way in which the gauge field ${A'}^\mu$ couples to the nucleus.  In particular when the field coupling to the nucleus is the photon, via kinetic mixing with a dark photon, the anapole and dipole operators give rise to an effective interaction of the form
\beq
\lagint^\text{Anapole}= {f_a \over  M^2-q^2} \chibar \gamma^\mu \gamma^5 \chi \sum_{N=n,p} \bar{N} \left(F_1^N \gamma_\mu +  F_2^N {i \sigma_{\mu \nu} q^\nu \over 2 m_N} \right)N
\eeq
\beq \label{magdip lint}
\lagint^\text{Magnetic Dipole}={f_d \over M^2-q^2} \chibar {i \sigma^{\mu \nu}  q_\nu \over \Lambda} \chi \sum_{N=n,p} \bar{N} \left(F_1^N \gamma_\mu +  F_2^N {i \sigma_{\mu \rho} q^\rho \over 2 m_N} \right)N
\eeq
where $F_i^N$ are the appropriate electromagnetic form factors, $M$ is the mediator mass and $q$ is four-momentum transfer \eqref{eq; mag dipole op}.\footnote{Note that the effective higher-dimension operator $\chibar \gamma^\mu \gamma^5 \chi \partial^\nu F_{\mu \nu}$, which is sometimes also referred to as ``the anapole operator'', may also be generated \cite{Ho:2012bg}.} The magnetic dipole operator in particular was shown in previous work to alleviate the tension between the constraints from the XENON100 experiment and the putative signals \cite{Fitzpatrick:2010br}. 

The goal of the present  paper is to re-examine the parameter space for light elastically scattering DM in light of the recent results from LUX, and the earlier constraints released from XENON100, XENON10 S2 only, CDMS Ge Low-Energy, COUPP, and PICASSO. We overlay CDMSlite bounds for the standard spin-independent case for reference.  CDMSlite data is not yet available to responsibly adopt their constraints to the operators considered in this paper. We do not include constraints from experiments such as Edelweiss \cite{Armengaud:2012pfa} and TEXONO \cite{Li:2013fla} whose limits are comparable to or surpassed in the low-mass region we consider by other experiments with the same nuclear target and similar low-energy thresholds (the Ge-target CDMS II low-energy analysis for these examples).  We also consider the constraints on isospin violating models \cite{Chang:2009yt,Feng:2011vu}, which modify the relative couplings to neutrons and protons to tune away the coupling to Xenon. We limit our attention to \emph{elastic} scattering of WIMPs off of nuclei; analysis of models with inelastic scattering is beyond the scope of this paper.   

We find that, unsurprisingly, LUX rules out the CDMS Silicon and CoGeNT regions of interest for all of the underlying WIMP-nucleon interactions we consider. If a more conservative choice for the nuclear recoil energy conversion is taken for the Xenon experiments, we find that small portions of the CDMS Silicon and CoGeNT regions of interest can survive the Xenon constraints, though typically in those regions other constraints enter that close the window. Under a more conservative assumption on nuclear recoil energy calibration, XENON100 and LUX constraints can be shifted up by 1-2 GeV in the $m_\text{DM}\sim 6$-$10$ GeV range. Of the models we consider, Anapole and Magnetic Dipole interactions do the best job of bringing the DAMA (assuming scattering primarily off of Sodium with quenching factor $Q_\text{Na}=0.3$), CoGeNT, and CDMS Silicon regions of interest into alignment: the three regions significantly overlap for anapole interactions and come close for the dipole. Even with the conservative assumption about nuclear recoil energy calibration, LUX still rules out the region where all three overlap for the Anapole interaction; PICASSO, XENON10 S2 and CDMS Ge low-energy are also competitive in this range. In addition to considering alternative energy calibration assumptions, we consider alternative halo models. The alternative assumptions we consider do little to weaken the LUX constraint relative to the CDMS Si and CoGeNT regions of interest. 

The outline of this paper is as follows.  In \S\ref{sec: rates} we specify nuclear scattering cross-sections for models we consider.  Then in \S\ref{sec: xenon constraints} we extract constraints for these models in parallel with the light DM CDMS Si, CoGeNT, and DAMA regions of interest. For the Xenon target experiments, we discuss and implement a very conservative alternative extrapolation of nuclear recoil energies. In \S\ref{sec: astrophysics}, we briefly discuss the effect of the halo model on our results. We conclude in \S\ref{sec: conclusions}.   

\section{Rates and Conventions for Light Momentum Dependent Dark Matter \label{sec: rates}}

We briefly review scattering rates to define our conventions.  The details of how we have derived the constraint or preferred region for each experiment are given in the appendix.  The rate for scattering is
\beq
\frac{d R}{d E_R} = N_T \frac{\rho_\text{DM}}{m_{\rm DM}} \int_{|\vec{v} | > v_\text{min}} d^3v v f(\vec{v},\vec{v}_e) \frac{d \sigma}{d E_R},
\eeq
where $v_\text{min} = \frac{\sqrt{2 m_N E_R}}{2 \mu_N}$ and $\mu_N$ is the DM-nucleus reduced mass.  To calculate rates we model the DM velocity distribution as a truncated Maxwellian distribution,
\beq
f(\vec{v}) \propto \left( e^{-(\vec{v}+\vec{v}_\text{e})^2/v_0^2} - e^{-v_\text{esc}^2/ v_0^2} \right) \Theta(v_\text{esc}^2 - (\vec{v}+\vec{v}_\text{e})^2), \label{eq: halo v distribution}
\eeq
where Earth's speed relative to the galactic halo is $v_\text{e} = v_\odot + v_\text{orb} \cos \gamma \cos[ \omega (t-t_0)]$, $v_0$ is mean WIMP speed relative to the galaxy, and $\vec{v}_\text{esc}$ is the galactic escape velocity.  We also use a standard value for DM density, $\rho_\text{DM}$.
Specifically, we take 
$
v_0=220~\text{km/s},~ 
v_\odot = 232~\text{km/s},~
v_\text{esc} = 544~\text{km/s},~
v_\text{orb} = 30~\text{km/s},~
\rho_\text{DM} = 0.3~\text{GeV}/c^2 / \text{cm}^3 ,~
\cos \gamma = 0.51, 
$ though as noted in \cite{Fitzpatrick:2010br}, modifying these parameters shifts the regions somewhat but does not alter the conclusions.  In the next section we consider in particular the effect of modifying the escape velocity on the constraints.

The differential rate is related to the scattering cross section off of a nucleus via 
\beq
\frac{d \sigma}{d E_R} = \frac{m_N \sigma_N}{2 \mu_N^2 v^2}.
\eeq
For the standard spin-independent case, this is related to the scattering off protons $\sigma_p$ via
\beq
\sigma_N^\text{SI} = \sigma_p \frac{\mu_N^2}{\mu_n^2}\frac{[f_p Z + f_n(A-Z)]^2}{f_p^2} F^2(E_R), 
\label{eq:spin-independent}
\eeq
where $\mu_n,~\mu_N$ are the nucleon-WIMP and nucleus-WIMP reduced masses, $f_p,~f_n$ are the proton and neutron couplings, $Z$ and $A$ are the atomic number and weight of the target nucleus, and we take the form factor $F(E_R)$ to be the Helm form factor.  For the standard spin-dependent case, we take
\beq
\sigma_N^\text{SD} = \sigma {\mu_N^2 \over \mu_n^2}{4 \over 3}{J+1 \over J}{(a_p \langle S_p \rangle + a_n \langle S_n \rangle)^2 \over (|a_p|+|a_n|)^2}
\label{eq:spin-dependent}
\eeq where $\sigma = (\sqrt{\sigma_p}+\sqrt{\sigma_n})^2$, with $\sigma_{p,n}$ the scattering cross-sections off protons and neutrons.  For $\langle S_{p,n} \rangle$ we take the values as in Table 1 of \cite{Cannoni:2012jq}.
  Here we are justified in neglecting the momentum-dependence of the spin-dependent nuclear form factor because we are specializing to the case of light ($m_{\rm DM} \lesssim 20 \text{GeV}$) DM where only small $|\vec{q}| b$, where $b$ is nuclear size, is relevant.  For the anapole and dipole cases, WIMPs couple to the electromagnetic current and lead to spin-independent, orbital-angular-momentum- and spin-dependent couplings.\footnote{See e.g. the appendix of \cite{Fitzpatrick:2012ib}.} The nuclear scattering cross-sections are
\begin{align}
\sigma_N^a  &= f_a^2{{\mu_{N}^2} \over \pi M^4} \left( Z^2 F^2(A;\vec{q}^2) \left( \vec{v}^2 - {\vec{q}^2 \over 4 \mu_N^2}\right) + {J+1 \over 3 J} {b_N^2 \over b_n^2} A^2 { \vec{q}^2 \over 2 m_N^2} \right) \label{eq: anapole rate}  \\
\sigma_N^d   &=  f_d^2{{\mu_{N}^2} \over \pi M^4} {\vec{q}^2 \over \Lambda^2} \left( Z^2 F^2(A;\vec{q}^2) \left(  \vec{v}^2 - {\vec{q}^2 \over 4 \mu_N^2} +{\vec{q}^2 \over 4 m_{\rm DM}^2} \right)  +  {J+1 \over 3 J} {b_N^2 \over b_n^2} A^2 { \vec{q}^2 \over 2 m_N^2}\right), \label{eq: magdip rate}
\end{align}
where $J$ is the spin of the nucleus, $b_N$ is the nucleus magnetic moment and $b_n=e/2 m_p$ is the nuclear magneton.
When reporting cross-sections, we use the convention $\tilde{\sigma} = f_a^2 \mu_n^2/\pi M^4$ for the anapole and $\tilde{\sigma} = f_d^2 \mu_n^2/\pi M^4$, $\Lambda=1$ GeV for the magnetic dipole. In addition, while recent work has suggested that the inclusion of proper nuclear responses may be important \cite{Fitzpatrick:2012ix,Anand:2013yka}, we have explicitly checked that, for the low momentum transfer relevant for light DM scattering, their momentum dependence is negligible.  Hence we proceed with only the usual spin-independent form factor.  For the $\vec{q}^2$ and $\vec{q}^4$ momentum dependent operators $\op_1-\op_3$, as done in \cite{Chang:2009yt} we will take the standard spin-independent  scattering cross-section in \eqref{eq:spin-independent} (for $\op_1$) or the spin-dependent scattering cross-section in \eqref{eq:spin-dependent}  (for $\op_2, \op_3$) and rescale it by a reference momentum-dependent factor, $(\vec{q}^2/\vec{q}_\text{ref}^2)^n$, where $n=1,~2$.  By default we take $|\vec{q}_\text{ref}|=1$ GeV. If the mediator mass is comparable to the momentum transfer, other important effects could occur, which we neglect here.

\section{Light Momentum Dependent Dark Matter versus Xenon constraints}\label{sec: xenon constraints}

Our results are shown in Figs.~\ref{fig: sig mDM plot AP}-\ref{fig: sig mDM plot SD} for spin-independent, anapole, dipole, spin-dependent, isospin-violating, and momentum-dependent DM.  We include fits to the CoGeNT \cite{Aalseth:2012if}, CDMS Silicon \cite{Agnese:2013rvf}, and DAMA results \cite{Bernabei:2010mq}, and constraints from the CDMS germanium low-energy analysis \cite{Ahmed:2010wy}, the XENON10 S2 only analysis \cite{Angle:2011th}, XENON100 \cite{Aprile:2012nq}, COUPP \cite{Behnke:2012ys} and PICASSO \cite{Archambault:2012pm}. 
The CDMSlite constraint of \cite{Agnese:2013lua} is shown on the spin-independent plot; we did not rescale the CDMSlite constraints for other forms of interactions because the collaboration has not yet released their data, and the shift in the constraint should mirror the shift in the CDMS germanium constraint.  

As is well known, the Xenon-target detector results are particularly sensitive to threshold effects and energy calibration issues, which we describe in more detail in the appendix.    Uncertainties in ${\cal L}_\text{eff}$ have been included in the constraint curves corresponding to XENON100 and LUX.  In the plots the dark blue and black curves correspond to the ${\cal L}_\text{eff}$ used by the XENON100 \cite{Aprile:2012nq} and LUX \cite{Akerib:2013tjd} collaborations, respectively, while the light blue and black curves correspond to a linear extrapolation of the average expected number of photo-electrons $\nu(E_R)={S_\text{nr} \over S_\text{ee}} L_y E_R {\cal L}_\text{eff}(E_R)$ for the $-1 \sigma$ boundaries of the ${\cal L}_\text{eff}$ measurement made by  Manzur et al.~\cite{PhysRevC.81.025808}. We effectively assume ${\cal L}_\text{eff}$ drops to zero at the lowest data point ($4$ keV). We show in Fig.~\ref{fig: leffs} in the appendix the ${\cal L}_\text{eff}$s we have used.  
For the S2 only XENON10 analysis, the dark red curve corresponds to the Lindhard model ionization yield $Q_y$ used by the collaboration in their analysis, and the light red curve corresponds to a variation of $Q_y$ as follows: We extrapolated from the $-1$ $\sigma$ boundaries of the ionization yield data points with $E_R>10$keVnr from the measurement by Manzur at $E_d=1$ kV/cm \cite{PhysRevC.81.025808}. We do a linear interpolation of $\{ \text{Log}E_R, Q_y \}$ including the point $Q_y(0)=0$.\footnote{More precisely, we set $Q_y(10^{-3} \text{keV})=10^{-3}e^-/\text{keV}$ for the log extrapolation.} We believe this is an appropriately conservative case to consider given that  (1) the Lindhard model is suspected to be a crude approximation at low energies for liquid Xenon, (2) there is significant disagreement between different measurements below about 10 keV, and (3) generic theoretical expectations are that the ionization yield should fall off at low energies.  For DAMA we have taken a quenching factor $Q_\text{Na} = 0.3$ for the most optimistic agreement with CoGeNT and CDMS-Si, though \cite{Collar:2013gu} suggests a lower $Q_\text{Na}\approx0.15$, which would shift the DAMA region to the right.  
For COUPP, we draw constraints under the two different assumptions about Fluorine efficiency adopted by the collaboration, as detailed in the appendix.  

Fig.~\ref{fig: sig mDM plot SI SD} establishes a baseline, showing all of the constraints we consider in the $\sigma$-$m_\text{DM}$ plane for isospin-conserving spin-independent and -dependent interactions. All constraints except for CDMSlite are derived independently, following the procedures laid out in the appendix. In subsequent plots we will show only an appropriately representative subset of constraints.

\begin{figure}
\includegraphics{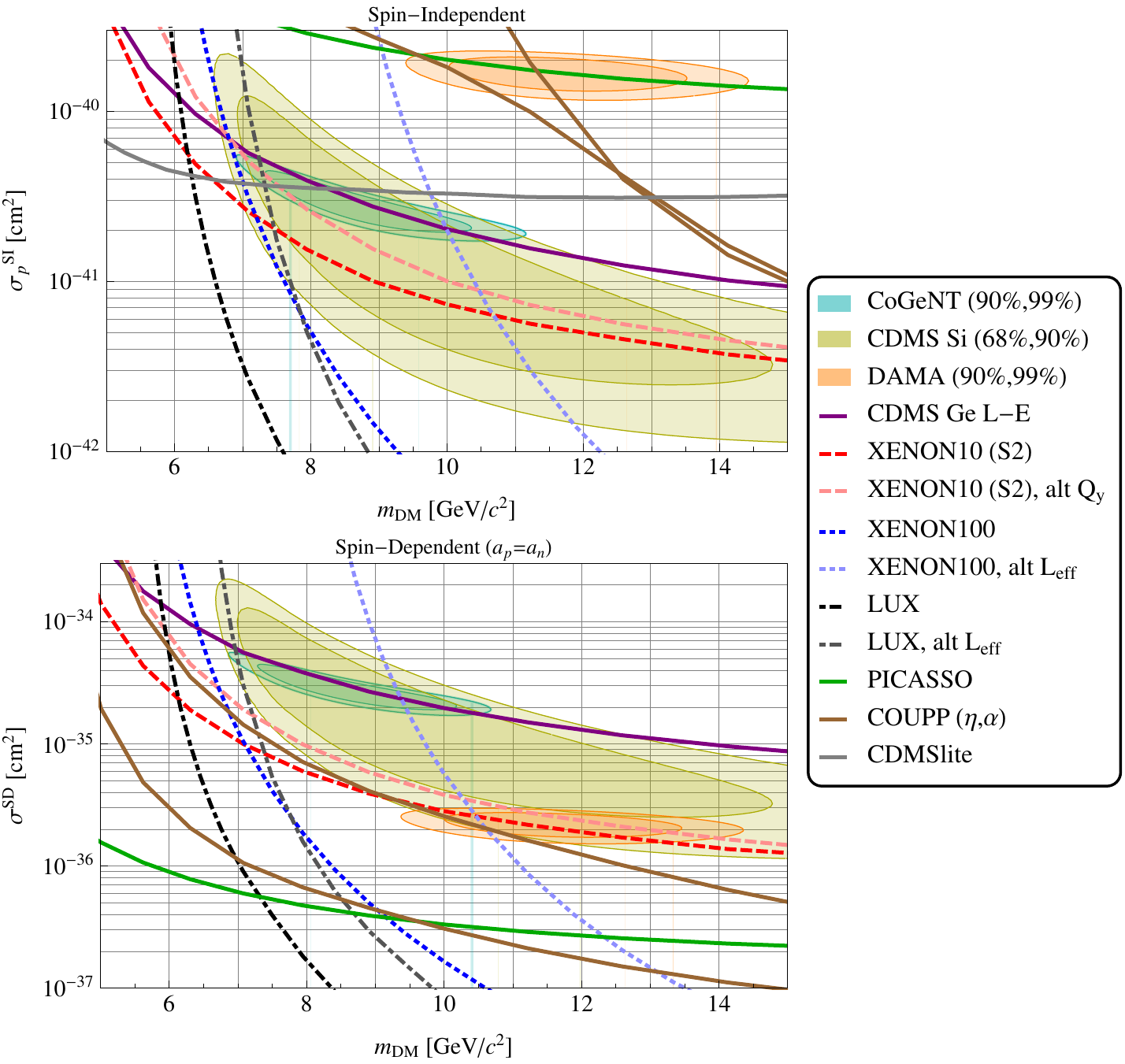}
\caption{Regions of interest and exclusion curves for experiments and parameters as listed in Table \ref{summary experiments}, assuming a  standard, spin-independent \eqref{eq:spin-independent} or -dependent \eqref{eq:spin-dependent}, isospin-conserving Nucleon-WIMP interaction. A standard Maxwellian distribution is assumed, as explained in the text. All constraint curves are 90\% C.L.~as explained in the appendix. We overlay the CDMSlite bound for reference; all other curves were generated as described in the appendix. We show both a weak and strong COUPP bound, as described in the appendix, and the choice of alternative ${\cal L}_\text{eff}$ for the Xenon experiments is shown in Fig.~\ref{fig: leffs} in the appendix.}\label{fig: sig mDM plot SI SD}
\end{figure}

\begin{figure}
\includegraphics{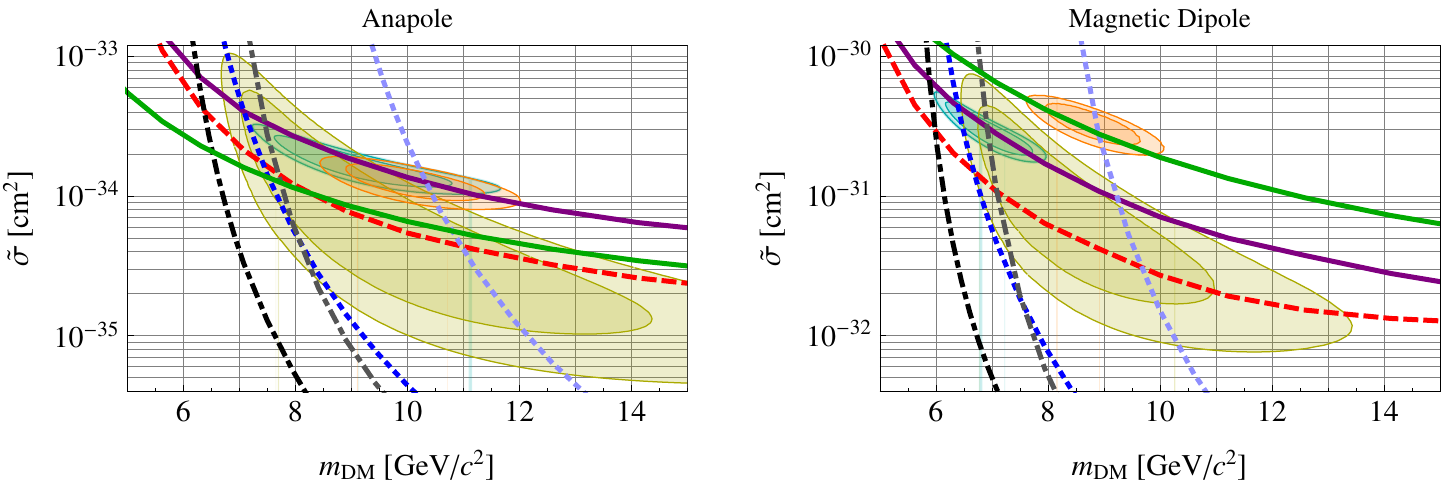}
\caption{Regions of interest and exclusion curves for relevant experiments and parameters as listed in Table \ref{summary experiments}, assuming an anapole \eqref{eq: anapole rate} or magnetic dipole \eqref{eq: magdip rate} Nucleon-WIMP interaction. We checked that the strong COUPP bound is weaker than the combination of LUX + PICASSO. Refer to Fig.~\ref{fig: sig mDM plot SI SD}.}\label{fig: sig mDM plot AP}
\end{figure}

\begin{figure}
\includegraphics{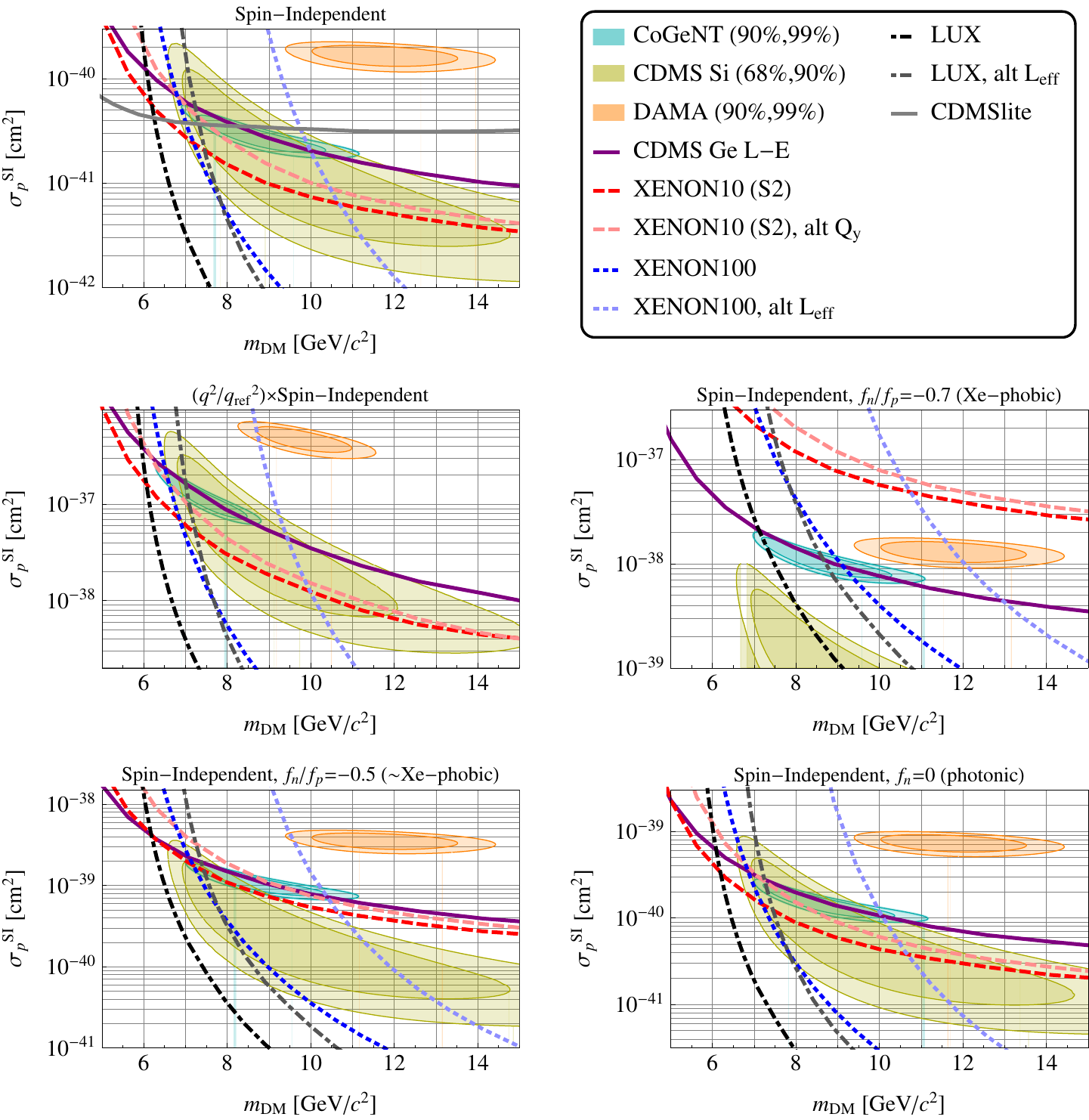}
\caption{ Regions of interest and exclusion curves for relevant experiments and parameters as listed in Table \ref{summary experiments}, assuming spin-independent Nucleon-WIMP interactions. We include constraints for a momentum-suppressed interaction (with $q_\text{ref}=1$ GeV) arising from scalar exchange as well as for some Xenonphobic isospin benchmarks in addition to the ``standard'' isospin-conserving case.}\label{fig: sig mDM plot SI}
\end{figure}

\begin{figure}
\includegraphics{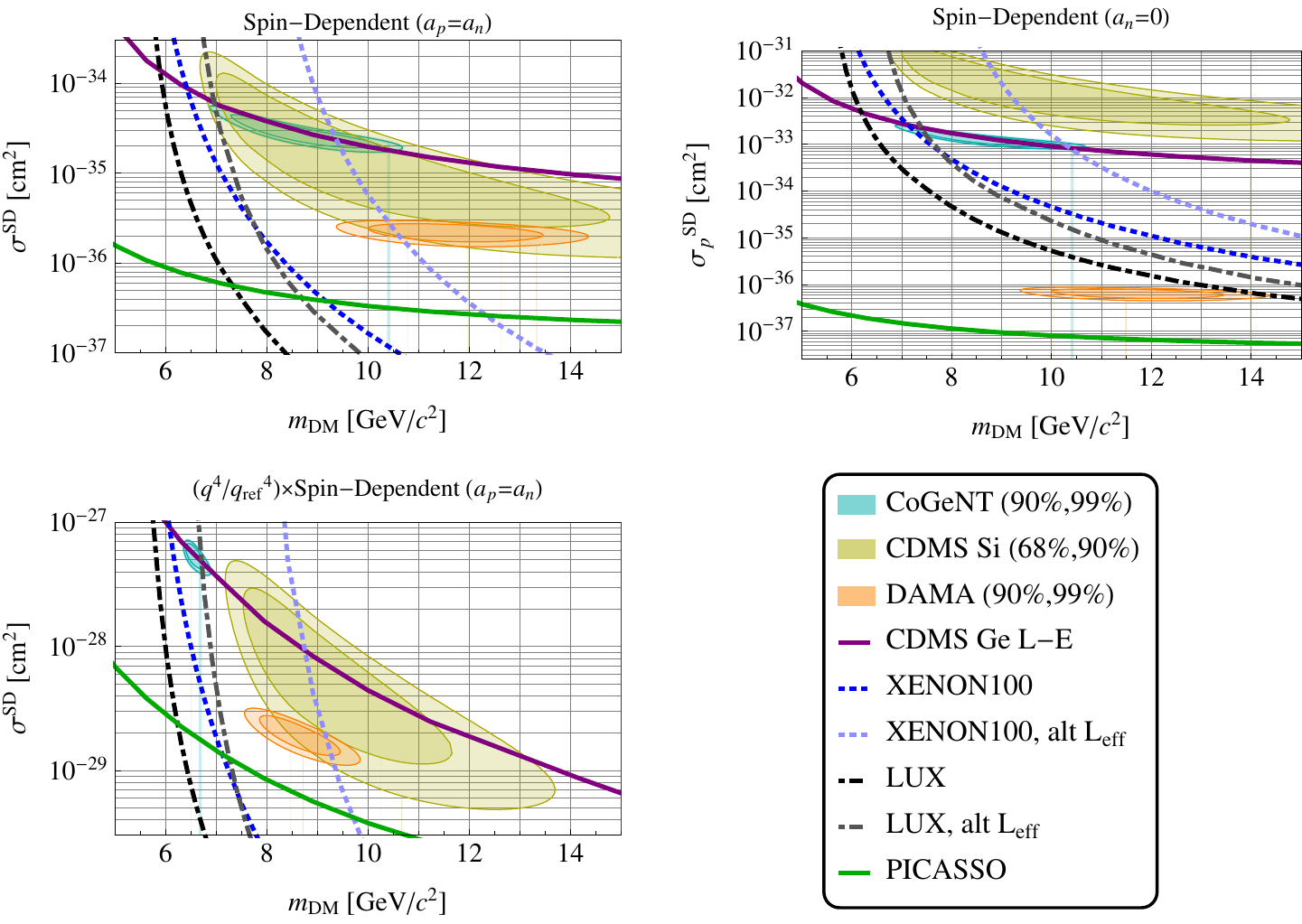}
\caption{ Regions of interest and exclusion curves for relevant experiments and parameters as listed in Table \ref{summary experiments}, assuming momentum-suppressed spin-dependent Nucleon-WIMP interaction arising from scalar exchange. We checked that the strong COUPP bound is weaker than the combination of LUX + PICASSO.}\label{fig: sig mDM plot SD}
\end{figure}

Fig.~\ref{fig: sig mDM plot AP} shows constraints and regions of interest for DM interacting via the anapole \eqref{eq: anapole rate} and magnetic dipole \eqref{eq: magdip rate} interactions, in the $\sigma$-$m_\text{DM}$ plane.  Since both the anapole and dipole have spin- and orbital-angular-momentum- dependent scattering components, we include the constraint from PICASSO as well, and we checked that the COUPP constraint is weaker than the PICASSO+XENON  bound throughout the region.  Both the anapole and dipole interactions bring the three regions of interest into good or marginal agreement, but the Xenon bounds do not loosen for the anapole in the region of interest relative to the spin-independent case. For the magnetic dipole, more of the CoGeNT preferred region is consistent with the LUX bounds, while remaining constrained by XENON10 S2 only. 

Fig.~\ref{fig: sig mDM plot SI} shows constraints and regions of interest for other spin-independent interactions, including momentum-suppressed interactions arising from Eq.~\eqref{eq: q2SI op} and isospin-violating interactions (see \cite{Feng:2011vu,Feng:2013vaa}), Eq.~\eqref{eq:spin-independent} with $f_n \neq f_p$.  Even given the ``xenophobic'' choice, $f_n/f_p = -0.7$, which minimizes DM coupling to Xenon, LUX still rules out all of the DAMA and most of the CoGeNT regions of interest, and much of the CDMS Silicon region of interest. Furthermore, while older studies emphasized that the xenophobic isospin choice brings the CoGeNT and DAMA regions of interest into ``agreement'', we can see that the 99\% C.L. regions for CoGeNT and DAMA are much closer than in the isospin-conserving case, but do not overlap with each other or with the CDMS Silicon region of interest. For the momentum-suppressed spin-independent interaction, the regions of interest shift towards lower masses to compensate for the momentum suppression, while XENON100 and LUX constraints shift relatively less since the larger target mass implies a larger momentum transfer in the scattering at a given nuclear recoil energy.  
This shift is not enough, however, to bring LUX into agreement with even the 99\% C.L. boundary of the CoGeNT region. Taking the very conservative choice for ${\cal L}_\text{eff}$ that we discuss opens up a corner of an overlapping CDMS/CoGeNT region of interest.

Fig.~\ref{fig: sig mDM plot SD} includes constraints for spin-dependent interactions \eqref{eq:spin-dependent}, including the most extreme momentum-suppressed interactions arising from \eqref{eq: q4SD op} and a couple of different choices for relative DM coupling to neutrons and protons.  The LUX and XENON100 bounds are very constraining even for spin-dependent interactions, regardless of whether the interactions are momentum-suppressed. We find a similar shifting of bounds and regions of interest in the momentum-suppressed cases as in \cite{Chang:2009yt}. The spin-dependent bound from  PICASSO 2012 is obviously very strong; it shifts relative to DAMA in the momentum-suppressed case, but not enough to bring the results into agreement. 

To summarize, using the rather conservative assumption for ${\cal L}_\text{eff}$ discussed above and in the appendix significantly loosens the XENON100 bound in all cases---enough to open up significant portions of the CDMS and CoGeNT (and in some cases, the DAMA) regions of interest. The LUX bound is also loosened, yet still strongly constrains most of the CDMS and CoGeNT regions of interest (and all of the DAMA region of interest except in the case of $a_n=0$ spin-dependent interactions) in all of the cases we consider. Unless ${\cal L}_\text{eff}$ or another aspect of deducing expected rates at LUX is severely misunderstood, and/or some alternative astrophysics is playing a dramatic role, a light DM elastic scattering explanation for the DAMA, CoGeNT and CDMS-Si anomalies appears to be substantially obstructed. 

In all cases one should keep in mind that the quenching factor $Q_\text{Na}=0.3$ we used for setting the DAMA regions has recently been claimed to bee too high especially for very low-energy recoils \cite{Collar:2013gu}. As noted above, lowering the quenching factor moves the preferred DAMA regions to higher masses---into worse agreement with CoGeNT and CDMS Si regions of interest.

\section{Astrophysical Dependence of Light Momentum Dependent Dark Matter Parameter Space}\label{sec: astrophysics}

\begin{figure}
\includegraphics{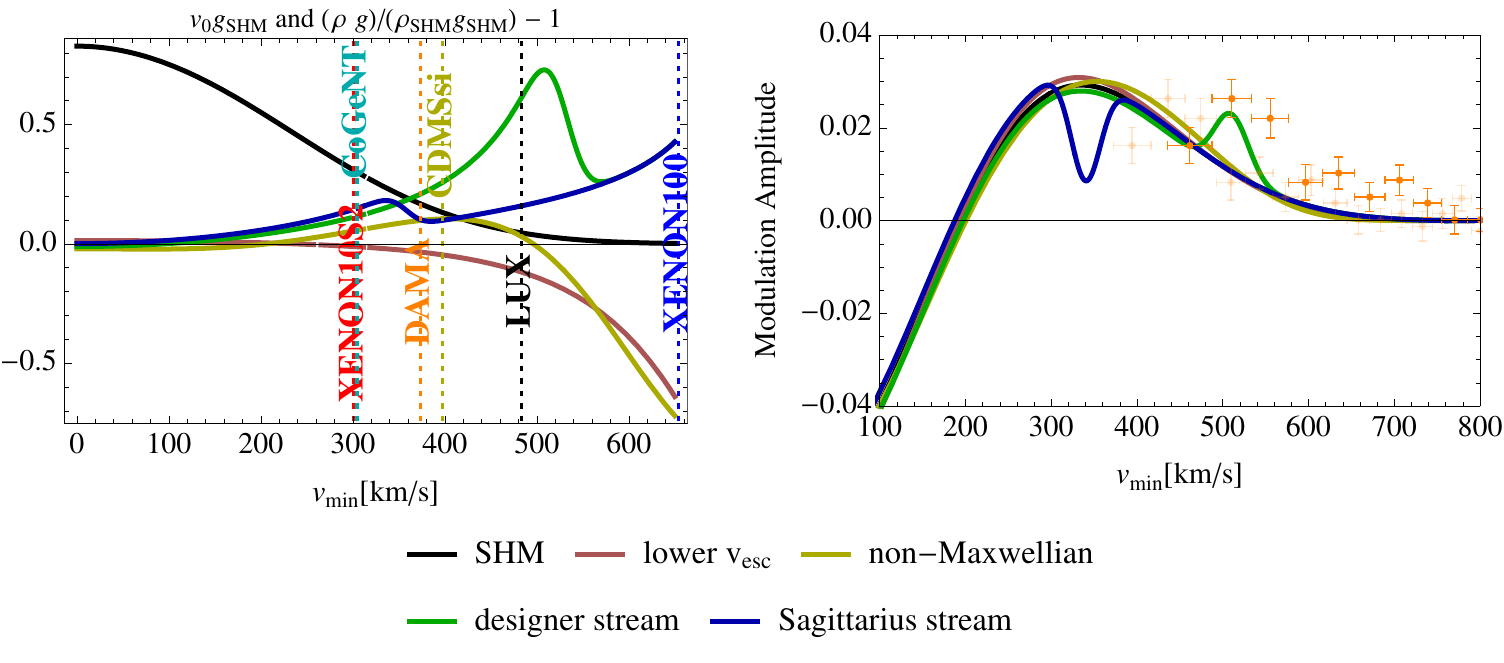}
\caption{\emph{Left:} velocity moment, $g_\text{SHM}(v_\text{min})$, for our standard halo model alongside the fractional difference ${g_\text{non-standard}\over g_\text{SHM}}-1$ for the ``non-standard'' distributions we consider, as indicated by the legend. Also shown is a dotted line at $v_\text{min} = \sqrt{2 m_N E_R^\text{min}}/2\mu_N$ for $m_{\rm DM}=10$GeV where $E_R^\text{min}$ is equal to the average expectation for nuclear recoil energy at the low end of the signal range for a given experiment. In other words, the dotted lines sit at the approximate minimum $v_\text{min}$ probed for $m_{\rm DM}=10$ GeV. For smaller (larger) $m_{\rm DM}$, the lines shift right (left). \emph{Right:} Annual modulation difference, $g|_\text{June 2} - g|_{\text{Dec 1}}$ relevant for modulation amplitudes for several different halo models as indicated by the legend.  Each stream is assumed to have density 5\% of the standard halo distribution. The ``designer stream'' is assumed to be in phase with the SHM and has characteristic velocities chosen to match the DAMA spectrum for $m_\text{DM}=8$ GeV. Overlaid is the spectrum of modulated DAMA events as a function of $v_\text{min}$ assuming $m_\text{DM}=8$ GeV. The lighter points indicate the spectrum assuming $m_\text{DM}=10$ GeV.}\label{fig:g plots}
\end{figure}

To conclude our discussion about a light DM elastic scattering explanation for anomalies, we consider the effects of modest changes in the assumptions of the standard halo model.  Fig.~\ref{fig:g plots} indicates how the high-velocity tail of the assumed velocity distribution can significantly affect light DM. The figure shows the velocity moment $g(v_\text{min}) = \int_{v_\text{min}}^\infty {1 \over v^2} v f(\vec{v}) d^3 v$ given a standard distribution as in  Eq.~\eqref{eq: halo v distribution} with standard choices of velocity parameters as described in the text, alongside the fractional difference for $g$ assuming several other velocity distributions: given a smaller galactic escape velocity,  an additional ``stream'' component modeled on the Sagittarius stream discussed in \cite{Savage:2006qr}, a stream designed to increase the modulation amplitude for DAMA, and the non-Maxwellian distribution of \cite{Bhattacharjee:2012xm}. Except in the case of leading velocity dependence in $\sigma_N$ for, {\em e.g.}, the anapole and magnetic dipole interactions,\footnote{These interactions depend also on the moment, $h(v_\text{min}) = \int_{v_\text{min}}^\infty  v f(\vec{v}) d^3 v$ due to the leading velocity dependence in the interactions.} the differential rate as a function of recoil energy is proportional to $g(v_\text{min}(E_R))$, which contains all of the astrophysics dependence in the rate \cite{Fox:2010bz,Gondolo:2012rs}. For order $10$ GeV DM, $v_\text{min}$ at the lowest recoil energies probed by LUX and XENON100 sits at the tail of $g$, as indicated on the figure. Thus the predicted rates for XENON100 and LUX for light DM depend highly on the high-velocity tail of the velocity distribution. It stands to reason that cutting the tail off at lower $v_\text{min}$---{\em e.g.} by lowering $v_\text{esc}$---could weaken the LUX and XENON100 constraints for light DM.

\begin{figure}
\includegraphics{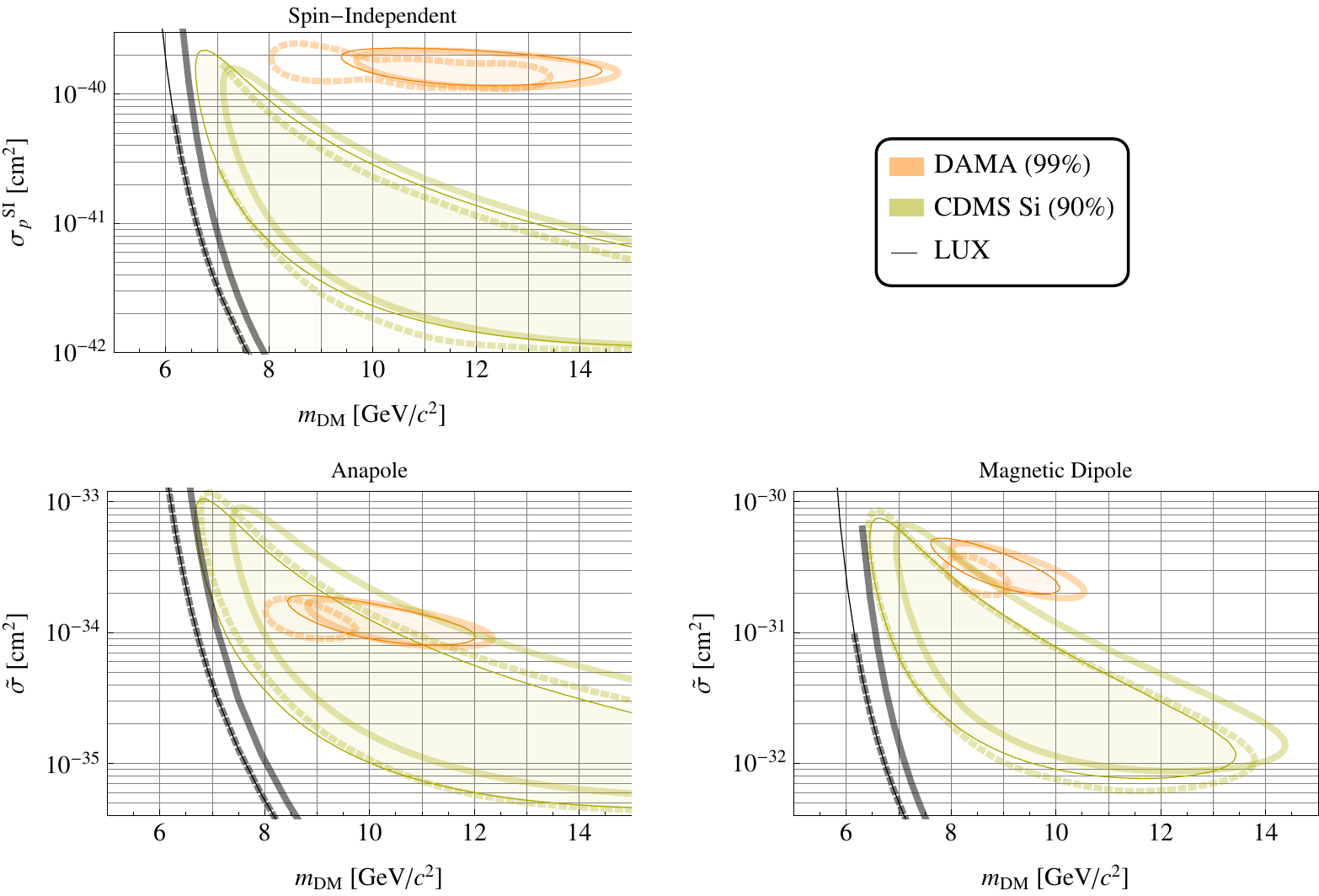}
\caption{ LUX bounds alongside regions of interest for CDMS Si and DAMA given (thin, darker) the standard truncated Maxwellian distribution with $v_\text{esc}=544$km/s, (thicker, lighter) a truncated Maxellian distribution with $v_\text{esc}=0.9\times544$ km/s, with $v_0=220$km/s and $v_\text{e}=232$ m/s fixed, and (thick, light, dotted) including a designer stream as in Fig.~\ref{fig:g plots}, modeled as an untruncated Maxwellian distribution with $v_\odot=510$km/s, $v_0=25$m/s, in phase with the primary distribution. We truncate the dotted designer stream LUX bound lines in two of the plots in order to reveal the overlap between the designer stream and SHM bounds for LUX.}\label{fig: alt halo}
\end{figure}

It has been noted that alternative halo distributions can affect modulation amplitudes quite dramatically while changing absolute rates very little \cite{Savage:2006qr}. This is because the modulation amplitude is sensitive to a different quantity: the change in $g$ at two opposite times of year. In Fig.~\ref{fig:g plots} we show this annual modulation difference assuming the alternate halo models discussed above. A stream with small dispersion can contribute a substantial peak even with modest density (in our examples, 5\%$\rho_\text{DM}$). Our ``designer stream'' is modeled as an untruncated Maxwellian distribution with $v_\odot=510$km/s, $v_0=25$m/s, and is in phase with the primary distribution. The DAMA modulation data points assuming $m_\text{DM}=8$ GeV (in order to convert to $v_\text{min}$) are overlaid on the $\Delta g$ plot to show that one can shift the DAMA preferred region towards a particular mass (in our example, 8 GeV) by tuning the velocity parameters of the stream. The light orange points show the spectrum for 10 GeV dark matter. Fig.~\ref{fig: alt halo} shows that the preferred DAMA regions shift much more dramatically (toward 8 GeV) than the Xenon constraint curves given our designer stream. 

We find that while reducing $v_\text{esc}$ to the marginally plausible value $v_\text{esc}=490$ km/s (see \cite{Smith:2006ym}) does weaken LUX and XENON100 bounds, it also moves the regions of interest so that increased agreement is not obtained.  This is shown in Fig.~\ref{fig: alt halo}. The cut-off of the high-velocity tail at lower velocity shifts the preferred regions towards higher masses at the same time that it weakens the Xenon constraint at a given mass. Since the Xenon bounds are nearly vertical in the relevant mass region, the weakening of the Xenon bounds does not win over the region-of-interest shift towards higher masses.

\section{Conclusions}\label{sec: conclusions}

We examined the parameter space for an elastically scattering light DM candidate to explain the DAMA, CoGeNT and CDMS-Si anomalies, through standard spin-independent and -dependent interactions, as well as anapole, dipole, and other momentum-dependent interactions.  In all cases, elastic scattering is in strong tension with the LUX results.  The tension is relaxed with a choice for the scintillation yield ${\cal L}_\text{eff}$ which is in the $-1 \sigma$ range as measured by Manzur et al \cite{PhysRevC.81.025808}, though most of the parameter space is still constrained. In particular, the anapole operator effectively brings all three anomaly-preferred regions into agreement, while not easing the constraints from the Xenon experiments; the dipole operator is most effective at reducing the tension with the Xenon constraints though not bringing the preferred regions of the anomalies into agreement.

We conclude that, absent a severe misunderstanding of experimental constraints at low recoil energy, the elastic DM scattering explanation of these anomalies is obstructed, and if a new physics explanation is to be found, more exotic types of scenarios should be sought.  At the same time, DM with mass below 10 GeV remains theoretically well-motivated ({\em e.g.} from models of Asymmetric DM and hidden sector models) and under-constrained in comparison to a 100 GeV DM candidate.  Thus further experimental investigation pushing to lower masses and smaller cross-sections continues to be warranted and compelling.

\acknowledgments
The work of KZ is supported by NASA astrophysics theory grant NNX11AI17G and by NSF CAREER award PHY 1049896.

\appendix

\section{Details for Event Rates and Experimental Constraints}\label{sec: experiments}

In this appendix we detail the analyses used in deriving bounds from various experiments. To translate from observed signal to a bound or region of interest, one must specify the expected number of events, event rate, or modulation amplitude in a given signal range, accounting for the resolution and cut efficiency of the experiment.
In general, the number of events expected in a given experiment signal range $[s_1,s_2]$ is
\beq
N_{[s_1,s_2]} = \text{Ex} \int_{s_1}^{s_2} \epsilon(s) \left( \int_0^\infty {dR \over dE_R} \mathcal{P}(E_R, s) dE_R \right) \, ds 
\eeq
where Ex is the exposure, $\epsilon$ is the efficiency, and $\mathcal{P}(E_R, s)$ is the probability per unit signal of observing signal $s$ given an actual recoil energy $E_R$.  For example, given perfect energy resolution and a mapping $\nu(E_R) = s$ from $E_R$ to $s$,  
\beq 
\mathcal{P}(E_R, s) = \delta\left(s-\nu(E_R)\right) ~~~ \text{so that} 
~~~N_{[s_1,s_2]} = \text{Ex} \int_{s_1}^{s_2} \epsilon(s) {dR \over dE_R} / \left( {d \nu \over dE_R}\right) \, ds. \eeq Depending on the experiment, the signal could be e.g.~electron equivalent energy ($E_\text{ee}$), ionization electrons (``S2''), or scintillation photo-electrons (``S1'').

The type of target, exposure, statistical method used in deriving bounds or regions of interest, reported signal type, analysis signal range, and total number of candidate events in the signal range are summarized for each experiment in Table \ref{summary experiments}. Below, we provide further analysis details. All constraint and region-of-interest curves using the procedures described below match well with those in the primary literature for spin-independent, isospin-conserving WIMP interactions.

\begin{table}
\begin{tabular}{l l l l l l l l l}
 			& $T$	& Ex 		& Stat.~Method   		& Ref. 				& \multicolumn{3}{l}{Signal \& keVnr Range}  & $N_\text{events}$ \\
\hline	
CDMS Si 		& Si 		& 140.2 kg-days 	& Max.~Likelihood 		& \cite{Agnese:2013rvf} 	& $E_R$ 			&				& 7-100	& 3 \\
DAMA 		& Na, I 	& 1.17 ton-yr 		& $\Delta \chi^2$  			& \cite{Bernabei:2010mq} &  $E_\text{ee}$	& 2-20 keVee\footnote{We used only data up to 14 keVee in our analysis.} & 6.7-67\footnote{For sodium, assuming a quenching factor $Q_\text{Na}=0.3$.}  \\
CoGeNT 		& Ge 	& 266 kg-days 		& $\Delta \chi^2$	& \cite{Aalseth:2012if} 	& $E_\text{ee}$ 		& 0.5-3 keVee		& 2.3-11	& 2272\footnote{Number after correcting for efficiency. Expected background $\sim$ 1640.} \\
CDMS Ge L-E 	& Ge 	& 35 kg-days\footnote{Used only data from detector T1Z5, which is the most constraining.} 
									& Yellin's $p_\text{max}$ 	& \cite{Ahmed:2010wy} 	& $E_R$ 		& 				& 2-100 	& 38 \\
Xenon10 S2 	& Xe 	& 15 kg-days 		& Yellin's $p_\text{max}$ 	& \cite{Angle:2011th} 	& S2  			& 5-43 ${e^-}$s 	& 1.4-10	& 23\\
XENON100 	& Xe 	& 7636 kg-days 	& Max.~Gap 			& \cite{Aprile:2012nq} 	& S1			& 3-20 PEs		&6.6-30.5	& 2 \\
LUX			& Xe		& 10065 kg-days 	& Max.~Gap			& \cite{Akerib:2013tjd}	& S1			& 2-30 PEs		& 3.6-24.8	 & 1	\\
PICASSO		& F		& 114.3 kg-days	& $\chi^2$			& \cite{Archambault:2012pm} &		& thresholds from:	& 1.7-55	\\
COUPP		& F,I		& 437.4 kg-days\footnote{After cuts.}	& 
									 Likelihood Ratio		& \cite{Behnke:2012ys} 	&			& thresholds from:	& 7.8-15.5	\\
\end{tabular}
\caption{Experiments/analyses considered in this work. We also include the target ($T$), total exposure (before cuts), statistical method used in setting bounds or regions of interest, the primary reference, the signal type reported ($E_R$ is recoil energy), signal range, recoil energy range (in keVnr),  and total number of candidate events passing all cuts. The nuclear recoil energy range quoted is the average expected energy corresponding to the signal range boundaries, so, generally speaking, energies on tails of distributions beyond this range are probed (though note the DM velocity distribution further limits the energy range probed).}\label{summary experiments}
\end{table}

\paragraph{CDMS Si}

We digitized the efficiency as a function of recoil energy shown in Fig.~1 of \cite{Agnese:2013rvf}. We approximated the resolution as being perfect. We maximized the log of the Likelihood function ($\ln L$) over DM mass and interaction strength, given the three candidate event energies and assuming zero background. Contours satisfying $\ln L = \ln L_\text{max} - \text{CDF}^{-1}(\text{ChiSq}[2], \text{C.L.}) / 2$ were drawn for C.L.=68\% and 90\%, where $\text{CDF}^{-1}(\text{ChiSq}[2], \text{C.L.})$ is the $\chi^2$ value at which the cumulative distribution function of a $\chi^2$ distribution for two degrees of freedom equals C.L..

\paragraph{DAMA}

We take the modulation amplitude to be 
\beq
\mathcal{A}(E_R)={1 \over 2} \left({dR \over dE_R}\bigg|_{v_\text{e}=v_\odot + v_\text{orb} \cos \gamma}-{dR \over dE_R}\bigg|_{v_\text{e}=v_\odot - v_\text{orb} \cos \gamma} \right)
\eeq
with $v_\text{orb} = 30$ km/s and $\cos \gamma = 0.51$. The expected modulation amplitude in energy bin $[E_1, E_2]$  is thus
\beq
S_{\text{m},[E_1,E_2]} = {1 \over E_2 - E_1} \sum_{T=\text{Na,I}} c_T\int_{E_1/Q_T}^{E_2/Q_T} \mathcal{A}_T(E_R) dE_R
\eeq
where $c_T$ is the mass fraction of the target, and $Q_T$ is the quenching factor for the target, which we take to be 0.3 for Sodium and 0.09 for Iodine.  A lower quenching factor for Sodium, as suggested by \cite{Collar:2013gu}, would worsen the agreement between DAMA and the results of CoGeNT and CDMS-Silicon in most cases.

We use the data in Fig.~6 of \cite{Bernabei:2010mq}. We calculate $\chi^2$ using the first 24 bins (bin widths are $0.5$keV) which corresponds to energies from $2$ keV to $14$ keV. The displayed 90\% and 99\% C.L.~region of interest contours satisfy $\chi^2 = \text{Min}(\chi^2) + \text{CDF}^{-1}(\text{ChiSq}[2], \text{C.L.})$. In some cases a DM focusing effect can be important for annual modulation experiments \cite{Lee:2013wza}, though for light DM scattering of Sodium, the effect is unimportant and we neglect it.  

\paragraph{CoGeNT}

We use the data in Fig.~23 of \cite{Aalseth:2012if}, which has been corrected for efficiency (i.e.~bin counts have been scaled to reflect the number of events expected based on those observed and the deduced efficiency). We do a $\chi^2$ scan over cross-section, DM mass, and a constant background component, using as errors those indicated by the error bars in the figure. Since correlations are not reported we are assuming that the bin-to-bin correlations are negligible. We then profile over the background. The region-of-interest curves correspond to the 90\% and 99\% C.L. regions.  More specifically, the contours are given by $\chi^2 = \text{Min}(\chi^2) + \text{CDF}^{-1}(\text{ChiSq}[3], \text{C.L.})$. Our understanding is that this is close to the procedure used by the collaboration.

The energy resolution below 10 keV is taken to be that reported by CoGeNT, namely
 $\sigma^2=\sigma_n^2+2.35^2 E \eta F$ where $\sigma_n =69.4$ eV is the intrinsic electronic noise, $E$ is the energy in eV, $\eta= 2.96$ eV is the average energy required to create an electron-hole pair in Ge at $\sim 80$ K, and $F=0.29$ is the Fano factor. The number of expected events in a given range is taken to be \beq
N_{[E_1,E_2]}=\text{Ex} \int_0^\infty  {dR \over dE_R} \text{res}(E_1,E_2;E_R)\,dE_R + b_{[E_1,E_2]}
\eeq
where $b$ is the flat, floating background and where $2 \, \text{res}(E_1,E_2;E_R)=\text{Erf}\left((E_1-E_R)/(\sqrt{2} \sigma)\right)-\text{Erf}\left((E_2-E_R)/(\sqrt{2} \sigma)\right)$.

\paragraph{CDMS Ge Low-Energy}

We used only the data from detector T1Z5, which contains the only events in the most constraining energy interval for 5-8 GeV DM \cite{Ahmed:2010wy}. The event energies and acceptance efficiencies for all detectors are provided as auxiliary files on the arXiv posting; we used the data in the file for detector T1Z5 for both the event energies and to extrapolate the efficiency. We assumed perfect resolution. 90\% C.L. limits were set using Yellin's $p_\text{max}$ method \cite{Yellin:2002xd}, which is very similar to Yellin's optimum interval method that was used in the CDMS analysis.

\paragraph{Xenon10 S2 only}

We use as input the highlighted candidate events shown in Fig.~2 of \cite{Angle:2011th}. An electron yield $Q_y = n_e/E_r$ as shown in Fig.~1 and given in Eq.~1 of \cite{Angle:2011th}\footnote{Note that $f_n=k g(E_R) / (1 + k g(E_R)$ where $g$ is the Lindhard function. We use the parameterization of $g$ found in e.g.~\cite{Lewin:1995rx} or \cite{2010PhRvC..81b5808M}. See also \cite{Sorensen:2011bd}.} was used by the collaboration.\footnote{Like the collaboration, we assume a sharp cutoff at $n_e=5$: $Q_y(E_R < E_R|_{n_e=5})=0$.} We analyze the data using both this electron yield
and an alternate electron yield as follows: we extrapolated from the $-1$ $\sigma$ boundaries of the ionization yield data points with $E_R>10$keVnr from the measurement by Manzur at $E_d=1$ kV/cm \cite{PhysRevC.81.025808}. We do a linear interpolation of $\{ \text{Log}E_R, Q_y \}$ including the point $Q_y(0)=0$.\footnote{More precisely, we set $Q_y(10^{-3} \text{keV})=10^{-3}e^-/\text{keV}$ for the log extrapolation.} We believe this is an appropriately conservative case to consider given the reasons explained in the text. 
A flat efficiency of 94\% was assumed. We also assumed an energy resolution $\sigma = E_R/\sqrt{E_R Q_y}$ so that 
\beq
N_{[E_1,E_2]}=\text{Ex} \int_0^\infty  {dR \over dE_R} \epsilon \, \text{res}(E_1,E_2;E_R)\,dE_R.
\eeq  We use the $p_\text{max}$ method of Yellin \cite{Yellin:2002xd} to set 90\% C.L.~exclusion curves. 

\paragraph{XENON100}\label{sec: xenon100}

\begin{figure}
\includegraphics{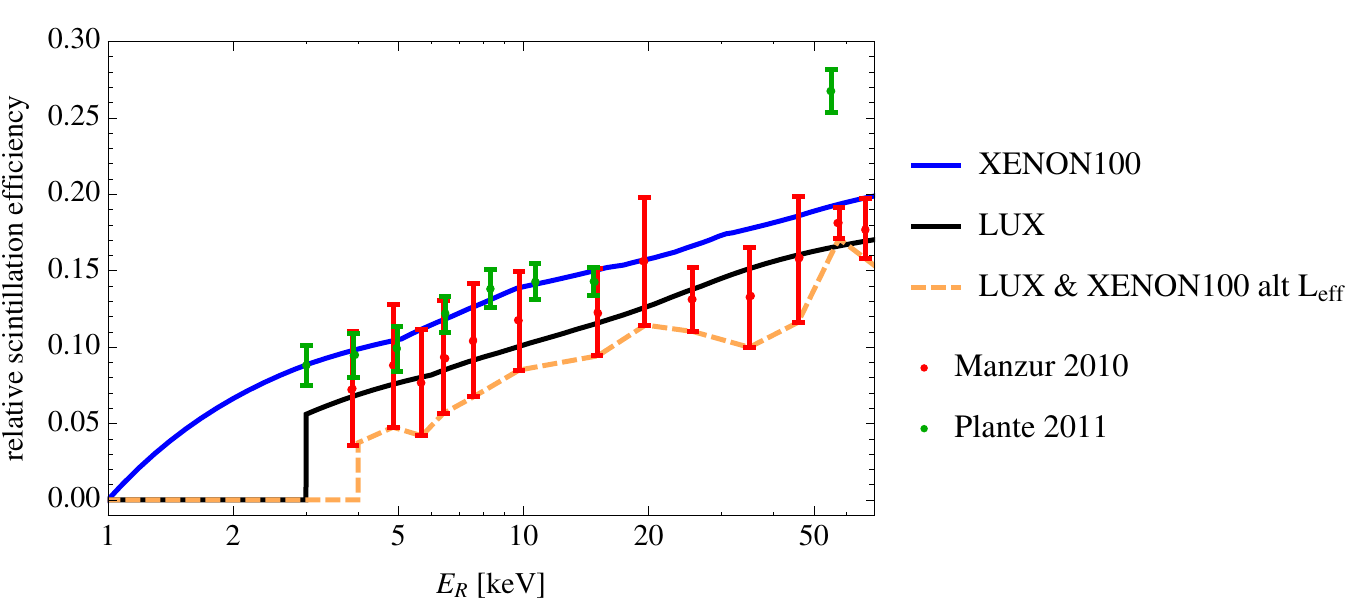}
\caption{ Relative scintillation efficiencies, ${\cal L}_\text{eff}$, used in this analysis. The ``alt ${\cal L}_\text{eff}$'' curve is based on the $-1\sigma$ boundaries of the measurement by Manzur et al \cite{PhysRevC.81.025808}. The main expected S1 function, $\nu(E_R)$ used in this analysis (blue for XENON100, black for LUX) was deduced directly from the scale on Fig. 1 of \cite{Aprile:2012nq} and Fig.~4 of \cite{Akerib:2013tjd} for XENON100 and LUX, respectively.}\label{fig: leffs}
\end{figure}

We digitize the efficiencies shown in Fig.~1 of \cite{Aprile:2012nq}, including the hard discrimination cut efficiency used for the maximum gap method analysis. The S2 threshold cut efficiency, $\epsilon_\text{S2}$, is applied ``before taking into account the S1 resolution'' \cite{Aprile:2012nq}. In addition to the red S2 threshold cut efficiency curve, the other efficiency curves from Fig.~1 (dotted green and blue) are digitized as functions of photo-electron  (PE) counts and are multiplied together to get $\epsilon$. Following \cite{Aprile:2012vw}, the number of events expected in signal range S1$\in[s_1,s_2]$ is taken to be
\beq
N_{[s_1,s_2]} = \text{Ex} \int_{s_1}^{s_2} \left[ \sum_{n=1}^\infty \epsilon(\text{S1}) \text{Gauss}(\text{S1}|n,\sqrt{n} \sigma_\text{PMT}) \int_0^\infty \text{Poiss}\left(n|\nu(E_R)\right) \epsilon_\text{S2}(E_R) {dR \over dE_R} dE_R  \right] d\text{S1} \label{eq: LXe expected}
\eeq
where $\nu(E_R)={S_\text{nr} \over S_\text{ee}} L_y E_R {\cal L}_\text{eff}(E_R)$ is the average expected number of photo-electrons if the nuclear recoil energy is $E_R$. Note that the S2 efficiency is set to zero below 1PE, corresponding to 3 keVnr, which is equivalent to setting ${\cal L}_\text{eff}$ to zero below $3$ keVnr. We use an interpolation of a digitization of the scale on Fig.~1 of \cite{Aprile:2012nq} for our default $\nu(E_R)$.\footnote{We also use the central values of the ${\cal L}_\text{eff}$ curve measured by the collaboration to check that we get similar $\nu(E_R)$.}  We use $\sigma_\text{PMT}=0.5$PE. To get an idea of the sensitivity of the XENON100 results on the energy calibration used, we also use a linear extrapolation of $\nu$ from the $-1 \sigma$ boundaries of the measurement of ${\cal L}_\text{eff}$ by Manzur \cite{PhysRevC.81.025808}, as shown in Fig.~\ref{fig: leffs}. To convert from ${\cal L}_\text{eff}$ to $\nu(E)$ we use the same values as XENON100: $S_\text{ee}=0.58$, $S_\text{nr}=0.95$, and $L_y=2.28$.  We read off the S2 cut efficiency as a function of S1 and take $\epsilon_\text{S2}(E_R)=\epsilon_\text{S2}(\nu(E_\text{R}))$; given the alternative ${\cal L}_\text{eff}$, 1PE corresponds to about 5.9 keVnr, meaning the alternative ${\cal L}_\text{eff}$ for XENON100 is effectively set to zero below 5.9 keVnr. We use the maximum gap method \cite{Yellin:2002xd} for the signal range $\text{S1}=3$PE to $\text{S1}=20$PE in order to set limits. Two events passed all acceptance cuts in this range.

\paragraph{LUX}

Instead of using a profile likelihood ratio test statistic like the collaboration, which also includes an expected background model and signal models in S1 and S2 from full simulations at each WIMP mass, and takes into account expected radius, depth, S1 and S2 of each event in a signal region that includes  regions with primarily electron recoils, we perform a simple maximum gap analysis much like the cross-check analysis used by XENON100.  We consider only events near or below the mean of the gaussian fit to simulated WIMP nuclear recoil (NR) data in slices of S1, in the S1-vs-$\log (\text{S2}_\text{b}/\text{S1})$ plane (the solid red curve of Fig. 4 of \cite{Akerib:2013tjd}).  We take the acceptance of this hard cut as a function of S1 to be 50\%. It is clear from Fig.~3 of \cite{Akerib:2013tjd} that this cut removes most electron recoil events. Only one event at $\{  \text{S1},\log (\text{S2}_b/\text{S1}) \} = \{ 3.2, 1.75 \}$ marginally makes the hard cut at the mean. We then set 90\% C.L. contours using a maximum gap analysis for the signal region of 2 to 30 PEs.

The LUX collaboration estimated the systematic uncertainty in the location of the NR band ``by averaging the difference between the centroids of simulated and observed AmBe data in $\log (\text{S2}_b/\text{S1})$'', which yielded an uncertainty of 0.044 in the centroid. If the centroid were moved up by 5\% in $\log (\text{S2}_b/\text{S1})$, the cut at the mean would still include only the same one event as if the centroid is as depicted in Fig.~4. If the centroid were lower, the one event should not make the cut.

Underlying particle- and astro-physics determines how signal events would be distributed as a function of energy. Since contours of constant NR energy are not vertical in the $\text{S1}$-$\log (\text{S2}_b/\text{S1})$ plane, the 50\% acceptance contour in the $\text{S1}$-$\log (\text{S2}_b/\text{S1})$ plane can slightly shift given different underlying WIMP physics. We note that for low NR energies relevant for light WIMP scattering, the constant energy contours are not far from vertical, so 50\% should remain a reasonable estimate of the acceptance  for light WIMPs given a cut at the centroid of the LUX simulated NR band. And again: given the actual distribution of events observed by LUX, a 5\% or so shift of the 50\% acceptance contour upwards (for any $\text{S1}$) does not affect the number of events making the cut. Below about 8 keV (note the recoil energy of xenon perturbed by 10 GeV DM moving at the galactic escape velocity is only 4.6 keV), even a shift of the 50\% acceptance contour all the way up to the $+1.28 \sigma$ LUX  NR contour would not add any events below 4.6 keV.  In other words, for light dark matter our procedure is robust to substantial shifts in the 50\% acceptance contour. Shifts downward would lead to our procedure being overly conservative. 

\begin{figure}
\includegraphics{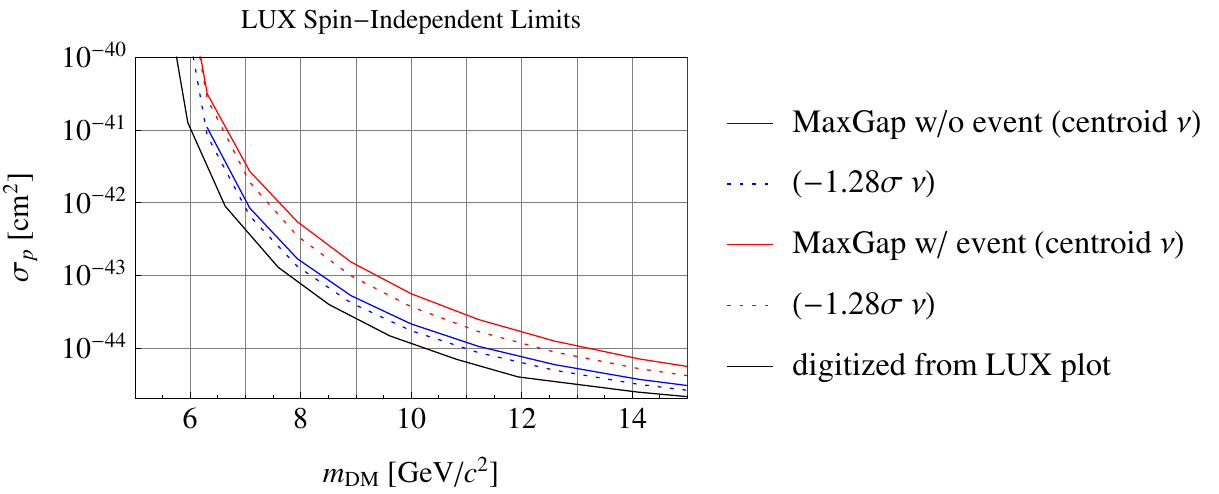}
\caption{Comparison of LUX bounds from maximum gap method described above. Red curves include the marginal event at $\{ S1, \log_{10} (S2/S1)\} = \{ 3.2, 1.75 \}$ while the blue curves do not. Dotted blue and red curves are drawn using $\nu(E_R)$ at the bottom of the NR band ($-1.28 \sigma$) while solid use $\nu(E_R)$ from the centroid of the NR band. See the text for discussion. The black curve is digitized from \cite{Akerib:2013tjd}.}\label{fig: lux compare}
\end{figure}

We take expected events to be as in \eqref{eq: LXe expected}; to do this we need $\nu(E_R)$, the mean expected $\text{S1}$ as a function of $E_R$. Since the collaboration models recoil energy as a function of both $\text{S1}$ and $\text{S2}$ (following \cite{Sorensen:2011bd}), we cannot directly follow the same procedure as LUX in converting to NR energy. However, for most NR events (in particular events falling within the NR band, within which energy does not vary much as a function of $\text{S1}$ over the relevant $\text{S2}$ range), energy can be reasonably reconstructed from just $\text{S1}$ \cite{Sorensen:2012ts}. We read $\nu(E_R)$ off of Fig.~4 of \cite{Akerib:2013tjd} by digitizing $\{S1, E_R\}$ values along the (red) centroid NR curve. We also deduce $\nu(E_R)$ along the  bottom  of the NR band ($-1.28 \sigma$) to get a feel for possible error introduced in making this choice. Taking $\nu(E_R)$ along the centroid is the more conservative choice (see Fig.~\ref{fig: lux compare}).
We interpolate the efficiency before the maximum gap cut from the NR simulation points (purple triangles) of Fig.~1 of \cite{Akerib:2013tjd}. The net efficiency is given by the efficiency from Fig.~1 of \cite{Akerib:2013tjd} times the 50\% for the maximum gap analysis cut.  The collaboration models no signal below 3 keVnr. We follow suit by taking $\epsilon_{S2}=\Theta(E_R-3 \text{keV})$ in \eqref{eq: LXe expected}. 

Fig.~\ref{fig: lux compare} shows that the bounds we get using the method described above are slightly weaker than, though close to, the bounds reported by LUX for spin-independent WIMPs, showing that our procedure is reasonable, and that our bounds are conservative compared to LUX's bounds. To get a feel for how much the bound can shift by adding (or subtracting) one observed event, we show curves generated assuming the one event at the NR band centroid does (red) and does not (blue) make the 50\% cut. Constraint curves drawn in all other figures in this paper were generated assuming the one event makes the cut. The dotted curves are drawn using $\nu(E_R)$ along the bottom (-$1.28 \sigma$) boundary of the NR band of Fig.~4. As expected, the constraint is not affected much by this small shift in $\nu(E_R)$. 

In order to more boldly estimate uncertainty due to the signal-energy conversion, as we did for XENON100, we use the alternative ${\cal L}_\text{eff}$ based on -1$\sigma$ boundaries of the Manzur measurement. We take an even more conservative line and cut ${\cal L}_\text{eff}$ off at the lowest measured point: $4$keV, as shown in Fig.~\ref{fig: leffs}. We deduced ${S_\text{nr} \over S_\text{ee}} L_y$ for LUX by comparing the $\nu$ read directly off of the LUX plots as discussed above to the ${\cal L}_\text{eff}$ function used by LUX \cite{lux-talk}. Again, see Fig.~\ref{fig: leffs}.

Though our procedure for drawing LUX bounds is necessarily less optimal than the procedure used by the LUX collaboration itself, we think it strikes a good balance between simplicity and sensible conservatism.

\paragraph{PICASSO}

PICASSO probes event rate as a function of recoil energy by relying on the fact that increasing the temperature of their liquid C$_4$F$_{10}$ target decreases the energy thresholds at which WIMP recoils can be detected. Therefore PICASSO is sensitive to integrated rates from a temperature-dependent threshold energy and above. Our PICASSO constraints are set by performing a simple $\chi^2$ fit to the integrated rates in Fig. 5 of \cite{Archambault:2012pm}; we read the eight rates and errors off of this plot. We take the expected rate as in Eq.~3, taking the resolution parameter to be $a=5$. Contours are set at $\chi^2 = \text{CDF}^{-1}(\text{ChiSq}[8],90\%)$. We emphasize that these constraint curves should only be expected to \emph{loosely} correspond to a 90\% C.L. limit. Our limit in the spin-dependent case is close to but slightly weaker than the limit shown in \cite{Archambault:2012pm}.

\paragraph{COUPP}

Like PICASSO, COUPP is sensitive to integrated rates above energy thresholds determined by the operating temperature of the bubble chamber liquid, CF$_3$I. We draw constraints based on their three different data sets, corresponding to three different bubble nucleation thresholds. We consider the events and expected backgrounds given the 530-sec time isolation cut that they discuss (last column of Table II) as we found this leads to better agreement with the curve in Fig.~6 (as did \cite{DelNobile:2013sia}); we assume that the acceptance for nuclear recoils above threshold is affected negligibly by this additional cut; we account for the overall 79.1\% efficiency to detect single bubble recoils after all of the other analysis cuts. We  consider scattering off of both Iodine and Fluorine.  We use two different assumptions for the efficiency of scattering off of Fluorine, like the collaboration: (a) that the efficiency turns on gradually, again following Eq.~3 of  \cite{Archambault:2012pm} but with the best-fit value of $a= 0.15$ and (b) that the scattering turns on abruptly at the threshold energy, but with 49\% efficiency. The efficiency for scattering off of Iodine is assumed to be 100\% above threshold. Assumption (b) leads to much stronger bounds on spin-dependent dark matter at low dark matter masses than assumption (a).
We use a log likelihood ratio statistic to draw 90\% C.L. contours according to $\sum_i-2 \ln \lambda_i = \text{CDF}^{-1}(\text{ChiSq}[1])$, where 
\[
\ln \lambda_i = N_i^\text{obs} \ln \left( N_{i}^\text{expected}(\sigma,m) + N_i^\text{bkgd} \over  N_i^\text{bkgd}\right) - N_{i}^\text{expected}(\sigma,m)
\]
and $i$ denotes the energy threshold bin.

\bibliography{direct_detection}

\end{document}